\documentstyle[psfig]{mn}

\def\simless{\mathbin{\lower 3pt\hbox
   {$\rlap{\raise 5pt\hbox{$\char'074$}}\mathchar"7218$}}}   
\def\simgreat{\mathbin{\lower 3pt\hbox
   {$\rlap{\raise 5pt\hbox{$\char'076$}}\mathchar"7218$}}}   
\def\etal{{\rm et al.}}

\def\solm{{M_\odot}}

\def\tff {t_{\rm ff}}

\def\Rc {R_{\rm clust}}
\def\ms {M_*}
\def\msi {M_i}

\def\Racc {R_{\rm sink}}

\def\vrel {v_{\rm rel}}
\def\vinf {v_{\rm inf}}
\def\macc {\dot M_*}
\def\menc {M_{\rm enc}}

\def\solm{{M_\odot}}
\def\be{\begin{equation}}
\def\ee{\end{equation}}
\def\macc{\dot M_*}
\def\racc{R_{\rm acc}}
\def\rbh{R_{\rm BH}}
\def\rroche{R_{\rm tidal}}
\def\rs{r_*}
\def\ms{M_*}

  \makeatletter
  \ifx\CUP@mtlplain@loaded\undefined  
  \makeatother  
    
  \else
    
  \fi

\makeatletter
\ifx\CUP@mtlplain@loaded\undefined  
  \newfont\bit{cmbxti10 at 9pt}
\else
  \newfont\bit{mtbxti10 at 9pt}
\fi
\makeatother  
 
\title[Accretion in stellar clusters] {Competitive accretion in embedded
stellar clusters} \author[Bonnell \etal] {I. A. Bonnell$^{1}$, M. R. Bate$^2$,
C. J. Clarke$^2$ and J. E. Pringle$^2$ \\ 
$^1$ School
of Physics and Astronomy, University of St Andrews, North Haugh, St
Andrews, Fife, KY16 9SS. \\
$^2$ Institute of Astronomy, Madingley
Road, Cambridge CB3 0HA.}

\date{\today}

\def\LaTeX{L\kern-.36em\raise.3ex\hbox{a}\kern-.15em
    T\kern-.1667em\lower.7ex\hbox{E}\kern-.125emX}

\begin{document}

\label{firstpage}

\maketitle

\begin{abstract}
We investigate the physics of gas accretion in young stellar
clusters. Accretion in clusters is a dynamic phenomenon as both the
stars and the gas respond to the same gravitational potential.
Accretion rates are highly non-uniform with stars nearer the centre of
the cluster, where gas densities are higher, accreting more than
others.  This competitive accretion naturally results in both
initial mass segregation and a spectrum of stellar masses.
Accretion in gas-dominated clusters is well modelled using a
tidal-lobe radius instead of the commonly used Bondi-Hoyle accretion
radius. This works as both the stellar and gas velocities are under
the influence of the same gravitational potential and are thus
comparable. The low relative velocity that results means that $\rroche
< \rbh$ in these systems. In contrast, when the stars dominate the
potential and are virialised, $\rbh < \rroche$ and Bondi-Hoyle accretion
is a better fit to the accretion rates.

\end{abstract}


\section{Introduction}

Star formation involves the collapse of molecular cloud cores under
their self-gravity through $\simgreat 20$ orders of magnitude in
density. This process is extremely non-homologous in that a small
fraction reaches stellar densities while the vast majority of the mass
is still infalling (Larson~1969). Thus, a stellar core, comprising a
small fraction of a solar mass, is initially formed which grows through
the accretion of the infalling envelope. It is the accretion which
ultimately determines the final stellar properties (Stahler, Shu \& Taam~1980; Palla~1999) as well as the observed characteristics during
its pre-main sequence evolution (Henricksen Andr\'e \& Bontemps~1997;
Andr\'e, Ward-Thompson \& Barsony~2000).

Complicating this picture is the fact that most stars do not form in
isolation but rather in groups from binary systems (Mathieu
\etal~2000) to stellar clusters (Clarke, Bonnell \& Hillenbrand~2000).
The separation between the individual components of these systems is
typically less than the size of an accreting envelope such that they
must compete for the reservoir of material.  

Surveys of star forming regions have found that the majority of
pre-main sequence stars (50 to 90 per cent depending on the region
considered) are found in clusters (e.g Lada et. al. 1991; Lada, Strom
\& Myers~1993).  These clusters contain anywhere from tens to
thousands of stars with typical numbers of around a hundred (Lada
et. al.~1991; Phelps \& Lada~1997; Clarke et. al.~2000).  Studies of
the stellar content of young (ages $\approx 10^6$ years) clusters
(e.g. Hillenbrand~1997) reveal that they contain both low and
high-mass stars in a similar proportion to that in a field-star
IMF (Hillenbrand~1997). Observations of young clusters have shown that
they are usually associated with massive clumps of molecular gas
(Lada~1992). Typically, the mass of the gas in the youngest clusters
is greater than that in stars (Lada~1991), with up to 90 \% of the
cluster mass in the form of gas.  Gas can thus play a crucial role
in the evolution of the clusters. Furthermore, there is a degree of
mass segregation present in the clusters with the most massive stars
generally found in the cluster cores (Hillenbrand \& Hartmann~1998;
Carpenter \etal~1997). These systems are generally too young for
two-body relaxation to explain these observations implying that mass
segregation is an initial condition of stellar clusters (Bonnell \&
Davies~1998).

Previous studies have explored the physics of accretion in binary
systems (Bate \& Bonnell~1997; Bate~2000) and in small stellar
clusters (Bonnell \etal~1997).  Accretion in binary systems plays an
important role in determining the system's mass ratio and separation
(Bate~2000).  In the more chaotic environment of a small stellar
cluster, the accretion rates are highly non-uniform with a few stars
accreting much more than the rest. The accretion rates 
primarily depend on the star's position within the cluster such
that those nearest the centre have the highest accretion rates (Bonnell
\etal~1997). This competitive accretion results in a spectrum of stellar masses, from initially equal
masses, and as such is a promising
candidate for explaining the observed initial mass function (Zinnecker~1982; Bonnell~2000).

In this paper we are extending the work on smaller stellar clusters to larger
systems containing up to one hundred stars. The aim is to explore the
physics of the competitive accretion in order to get a better grasp of the
likely effects of competitive accretion on the spectrum of stellar masses.
In \S~2 we discuss the numerical calculations. In \S~3 we discuss how the
accretion is related to the cluster dynamics. Section~4 presents the
results on accretion and mass segregation while \S~5 presents our
results on modelling the physics of competitive accretion. Our
conclusions are presented in \S~6.

\section{Calculations}

The calculations presented here were performed with a hybrid SPH-Nbody
code to model the combined presence of stars and gas and their
mutual interactions. This code is based on a standard 3-D Smoothed
Particle Hydrodynamics (SPH) code (Benz~1990) with gravitational
interactions calculated via a tree-code (Benz \etal~1990). The gas is
modelled by the normal SPH particles while the stars are modelled
using sink-particles (Bate, Bonnell \& Price~1995).  These
sink-particles interact with other particles only though gravitational
forces and the accretion of the gaseous SPH particles. The accretion
is modelled by removing gas particles within a predetermined radius of
one of the sink-particles, providing they are bound to the
sink-particle, and adding their mass and momentum to the
sink-particle's (Bate \etal~1995). The radius at which this accretion
occurs, the sink-radius $\Racc$ is chosen to be sufficiently small
that it does not overtly affect the gas flow outside this radius. In
practice this means that it must be smaller than any physical
accretion radius (e.g. Bondi-Hoyle accretion radius) where the gas becomes
bound to the star. The sink-radius cannot be too small either as this
translates into a prohibitively small time-step. In this study, we
use a sink-radius of $\Racc = 10^{-3} \times \Rc$, the initial 
cluster radius.

The simulations were performed with $10^5$ SPH particles to model the gas. 
The end-point of
the simulations was generally chosen to be when roughly half of the
SPH particles were accreted, although several were run till all the
gas was accreted. The analysis was performed before the majority of
the gas was accreted in order to ensure sufficient
resolution. Furthermore, several simulations were rerun with accreted
particles each replaced by nine lower-mass particles to ensure that
the results were not overtly affected by the resolution. This was
indeed found to be the case in all the simulations except those with
both `hot' gas and stars where the low accretion rates require larger
particle numbers to be fully resolved.
Limitations of the simulations presented here are that there is
no feedback from the stars onto the gas and that the gas does not
initially contain any turbulent (or other) motions. These two
effects could play a significant role in the cluster evolution and in the 
accretion process but are beyond the scope of this paper. 
 
\subsection{Initial Conditions}

\begin{table*}
\begin{center}
\begin{tabular}{c|c|c|c|c|c|c} \hline
Run &No. of stars & Gas fraction & stellar dist$^{\rm n}$ & $N(M_J)$ & Mach no.& Initial masses\\
 \hline
Run A & 100 & 90 & uniform & 120 & 0.5 & equal \\
Run B & 100 & 90 & $r^{-2}$ & 2.8 & 0.2 & equal \\
Run C & 100 & 90 & uniform & 120 & 0.5 & range \\
Run D &1 & 99 & --- & 130 & 0.2 & --- \\
Run E &1 & 99 & --- & 130 & 2. & --- \\
Run F & 30 & 82 & uniform & 110 & 0.6 & equal \\
Run G & 100 & 90 & $r^{-2}$ & 120 & 0.5 & equal \\

\end{tabular}
\caption{\label{models} The initial cluster models used to illustrate the
physics of the accretion process are listed by the number of stars in the cluster, the percentage of total mass initially in gas, the stellar distribution,
the number of Jeans masses contained in gas, the mean Mach number of the stars
and the distribution of initial stellar masses.}
\end{center}
\end{table*}

We have simulated accretion onto a large number of cluster models to
cover the range in possible initial conditions.  We present the
results from a number of these simulations, listed in Table~\ref{models}, in
order to illustrate the relevant physical processes of accretion in
stellar clusters. The simulations are characterised by the number of
stars, the gas fraction, the stellar distribution and the number of
Jeans masses in the gas, 
\be
\label{Jeans_mass}
M_J = \left(\frac{5 R_g T}{2 G \mu}\right)^{3/2} \left(\frac{4}{3} 
\pi \rho \right)^{-1/2},
\ee
where $R_g$ is the gas constant, $G$ is the gravitational constant, and
$\mu$ is the mean molecular weight. Other cluster parameters are the initial mean Mach number of the stars and the distribution of stellar masses.

All the clusters considered here are initially spherical and have
uniform gas distribution. In all cases, the gas is initially static
and has only thermal support. The stars  initially  hav a small virial ratio
in all cases. The initial conditions are termed
`cold' when they contain at least as many Jeans masses as stars. Such
initial conditions are the most likely if the cluster forms through a
fragmentation event.   
An isothermal gas equation of state is
used throughout. This means that, in the absence of accretion, the gas
will collapse to a singularity in approximately one free-fall time.

The initial spatial distribution of stars is either
uniform or $n(t) \propto r^{-2}$ as might be expected after a violent
relaxation. One simulation (Run~C) was performed with an initial range
of stellar masses in order to ascertain the importance of the initial
stellar mass distribution. 

\section{Accretion and cluster dynamics}

One of the first things we need to understand in terms of the dynamics
of accretion in clusters is the timescale. This sets the mean
accretion rate and thus how important accretion can be in setting the
final stellar mass.
Gas accretion by a star is given by the general formula
\be 
\macc \approx \pi \rho \vrel \racc^2,
\ee
where $\rho$ is the gas density and $\vrel$ is the relative gas-star
velocity and $\racc$ is the accretion radius.

One possible estimate of the accretion timescale is 
using Bondi-Hoyle accretion (Bondi \&
Hoyle~1944; Bondi~1952), corresponding to  
an isolated star accreting from a uniform,
non-self-gravitating medium.
In this model, the accretion radius is
given by
\begin{equation}
\rbh = { C_{\rm BH} G\ms \over \vrel^2 + c_s^2}, 
\ee 
where $\ms$ is the
stellar mass, and $c_s$ is the gas sound speed and the constant $C_{\rm BH}$
is generally taken to be $C_{\rm BH}\approx 2$ (Bondi \& Hoyle~1944).  
This approach
neglects the self-gravity of the gas, the presence of other stars, the
cluster potential and how these affect the accretion.  Estimating
the accretion timescale as
\be
t_{acc} = {M_{\rm gas} \over N \macc}
\ee
the  Bondi-Hoyle formalism implies
\be
t_{acc}(BH) \approx {1\over N} \left({\Rc\over \rbh}\right)^2 t_{dyn}.
\ee
If the cluster is virialised 
\be
{\Rc\over \rbh} \approx N,
\ee
then
\begin{equation}
t_{acc}(BH) \approx N t_{dyn},
\end{equation}
where $N$ is the number of stars in a cluster and $t_{dyn}$ is the
cluster's dynamical or crossing time. In studies of  accretion in
small-N clusters (Bonnell \etal~1997), this timescale was found to be
much too long as it neglects the motion of the gas in the cluster.

Another possible estimate is obtained by considering how
the gas evolves under the cluster potential. 
Assuming the gas is gravitationally unstable (otherwise it would not have
formed stars in the first place), dominates the cluster potential, lacks rotational support and remains isothermal,
the gas falls towards the centre of the cluster on the
free-fall timescale,
\be
\tff = \sqrt{3 \pi \over 32 G \rho}.
\ee
In the absence of any accretion onto existing stars, the gas
collapses and forms an additional star containing all the gas mass 
in approximately one free-fall time. This sets an upper limit on the accretion 
timescale as the gas remains distributed in the cluster for no more
than approximately one free-fall time.

During the collapse, the
gas density increases and the cluster potential becomes deeper. Thus
as long as the stars remain inside the gas sphere, the increase in the
gas density will increase the accretion rate (equation~2). As the density
increases towards infinity on a dynamical timescale, the accretion rate also
increases on this timescale till all the gas is accreted (or has formed another star at the cluster centre). Thus, the accretion timescale in
a self-gravitating gas is given by 
\begin{equation}
t_{acc} \approx \tff.
\end{equation}

\begin{figure}
\vspace{-0.35truein}
\psfig{{figure=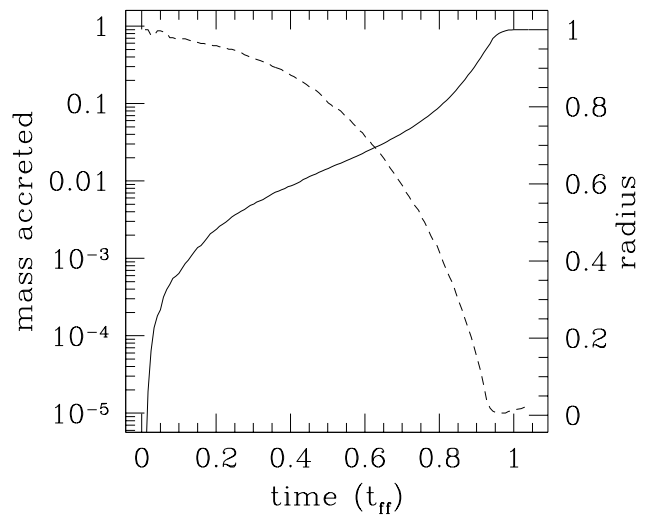,width=3.45truein,height=3.45truein}}
\caption{\label{accvstime} The total accreted mass (solid line) is
plotted as a function of time for Run~A (a 'cold' cluster containing 100
stars and 90 per cent of its initial mass in the form of gas). The
half-mass radius of the stars is also plotted (dashed line). Note
that both the gas accretion and the cluster evolution occur on a dynamical timescale.}
\end{figure}

Comparing these estimates to the numerical simulations, we find that
the accretion occurs on the dynamical timescale as the gas
density increases dramatically on this timescale.
Figure~\ref{accvstime} plots the evolution of the total
accreted mass versus time for Run~A (a cold cluster containing 100 stars
with an initial gas fraction of 90 per cent). The stars initially have a total
mass of 0.1. They accrete somewhat slowly over the first half of a free-fall
time and only double their mass by $0.8\ \tff$. By $0.95\ \tff$ the stars
have accreted all the gas in the cluster and have increased their mean
mass tenfold. 

The half-mass radius of the stellar distribution is also plotted in
Figure~\ref{accvstime} showing how the cluster evolves on a dynamical
timescale.  The stars are initially cold and thus in the absence of
any gas accretion would revirialise at half their initial
radius. Instead, the stars' dynamical evolution is due to the
combination of the accretion and the changing cluster potential due to
the collapse of the gas. As the gas collapses, the deepening of the
cluster potential forces the stars to follow the gas and
shrink. Additionally, the accretion increases the stellar masses
making them more bound to each other which causes the stellar
distribution to shrink further. Thus not only does the accretion occur
on a dynamical timescale but the cluster as a whole evolves on this
timescale. This evolution may be important in forming massive stars
if they cannot accrete, due to their large radiation pressure, past
$\approx 10 \solm$. In this case, the cluster shrinkage due to the accretion
can be sufficient to induce stellar collisions which form the most
massive stars (Bonnell, Bate \& Zinnecker~1998). Stellar collisions
are not included in the simulationss presented here as the finite stellar size is not modelled. The reader is referred
to Bonnell \etal~(1998) for a discussion on how accretion-induced shrinkage
of the clsuter can result in stellar collisions.

Although many of the clusters investigated here are cold and contain
$\approx 100$ Jeans masses, they do not in general collapse to form
additional stars. Instead, the gas was accreted onto individual stars
that the gas encountered on its way to the cluster centre. Exceptions
to this were the simulations run with only one star present in the gas
cloud. In these cases the gas can easily collapse to the centre
without encountering (being accreted by) the single star.

\section{Accretion and mass segregation}

One of the goals of studying accretion in stellar clusters is to determine
if it can explain the initial mass segregation that is observed in
young stellar clusters (e.g. Hillenbrand \& Hartmann~1998). This
segregation of the more massive stars towards the centre of the
cluster is not due to two-body relaxation in the cluster as the
clusters are often too young (Bonnell \& Davies~1998). 
Moreover, simple Jeans mass arguments imply  that the
lowest  mass stars should be found near the centre of the cluster
where the gas density is highest, not the most-massive
ones (Zinnecker, McCaughrean \& Wilking~1993; Bonnell \etal~1998). Simulations of accretion
in small stellar clusters (Bonnell \etal~1997) found that the
accretion rate was generally higher for the stars nearer the cluster
centre. In this section, we investigate how the accretion rate depends
on the stellar positions in the cluster and how this can result in a
mass-segregated cluster.

\begin{figure}
\vspace{-0.35truein}
\psfig{{figure=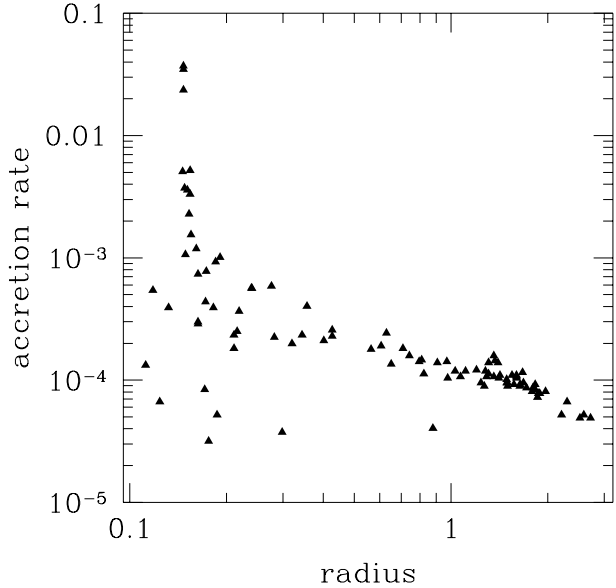,width=3.5truein,height=3.5truein}}
\caption{\label{accratevsrad} The accretion rate onto 100 stars in a
cluster is plotted as a function of the radius in the cluster of each
of the stars (Run~B). The accretion rates are averaged over $0.1\tff$
and are calculated at $t\approx 0.65\tff$.  The accretion rate
increases towards the centre of the cluster which leads to mass
segregation. Initially the cluster is centrally condensed and contains
10 per cent of its mass in stars. The accretion rate is in units of
the total cluster mass per $\tff$.}
\end{figure}

Accretion leads to mass segregation in two ways. Firstly, as the
system relaxes to a centrally-condensed distribution, the gas density
increases towards the centre of the cluster which then results in
higher accretion rates in the centre than elsewhere. Secondly, mass
accretion onto an individual star decreases its kinetic energy (mass
loading) which forces it to sink deeper into the cluster
potential. Thus, stars that accrete gas sink towards the centre and
stars that are in the centre accrete more gas as the density
is highest there.  

Figure~\ref{accratevsrad} plots the individual accretion rates versus
stellar position in the cluster for Run~B.  These accretion rates are
averaged over approximately $0.1 \tff$ at time $t \approx 0.65
\tff$. The 100 stars are initially centrally condensed in this model
and the gas, comprising 90 per cent of the total mass, is relatively
warm in that it contains only 2.8 Jeans masses. The stars dominate the
central part of the potential due to their condensed distribution.
Figure~\ref{accratevsrad} shows that the accretion rate increases with
decreasing radius with those stars near the cluster centre accreting
at a rate more than a hundred times those near the outside. This
differential accretion is due to the higher gas densities found near
the centre of the cluster and obviously leads to a mass-segregated
cluster on the accretion (dynamical) timescale. The most massive stars
are the ones that started out nearest the cluster centre. They have
the largest accretion rates and will therefore always be nearest the
cluster centre (unless ejected by a dynamical interaction).

\begin{figure}
\vspace{-0.35truein}
\psfig{{figure=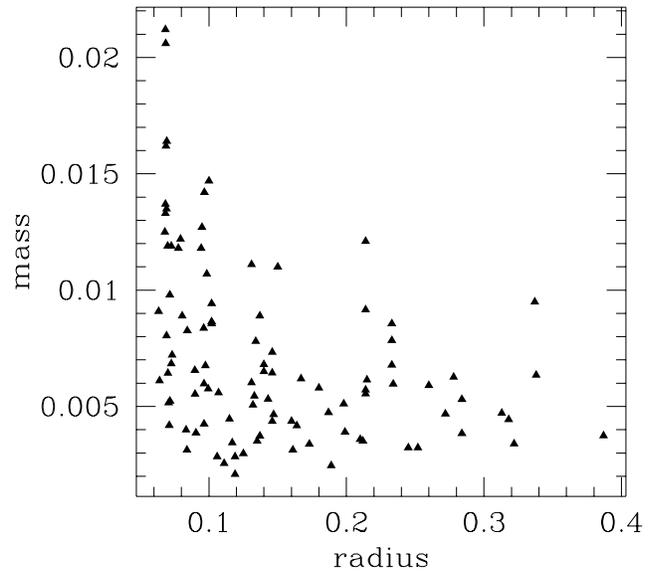,width=3.5truein,height=3.5truein}}
\caption{\label{massvsrad} The resultant stellar masses (in units of
the total cluster mass) after accretion are plotted as a function of
each star's position in the cluster at $t\approx 1\tff$ for
Run~A. The initial stellar mass is $m_{\star} = 0.001$. The mass
segregation is apparent even though there are significant numbers of
low-mass stars near the cluster centre. The cluster is initially
uniform and cold and contains 10 per cent of its mass in stars.}
\end{figure}

For clusters whose stars are initially uniformly spatioally distributed,  
the gas
density is initially uniform as well and only becomes significantly
centrally concentrated on a free-fall timescale. In this situation, it
is not necessarily the stars initially nearest the cluster centre which
accrete the most. Instead, it is those which, regardless of
how close they are to the cluster centre, are in a favourable
location (e.g. less local competition) that accrete the
most. These stars then sink towards the centre and continue to accrete
at a higher rate. This still results in a mass segregated cluster but
there is no direct correlation between the stars' {\it initial}\ positions in
the cluster and their eventual mass. An example of the resulting mass
distribution in the cluster is given in Figure~\ref{massvsrad} for Run~A (an
initially uniform cluster containing 100 stars with 90 per cent of its
mass in the form of cold gas and containing 120 Jeans masses). We note
here that although the trend of higher mass stars towards the centre
is very distinct, there are still low-mass stars near the cluster
centre as is observed in young stellar clusters such as the ONC
(Hillenbrand~1997; Hillenbrand \& Carpenter~2000).

\section{Accretion and stellar masses}

\begin{figure}
\vspace{-0.35truein}
\psfig{{figure=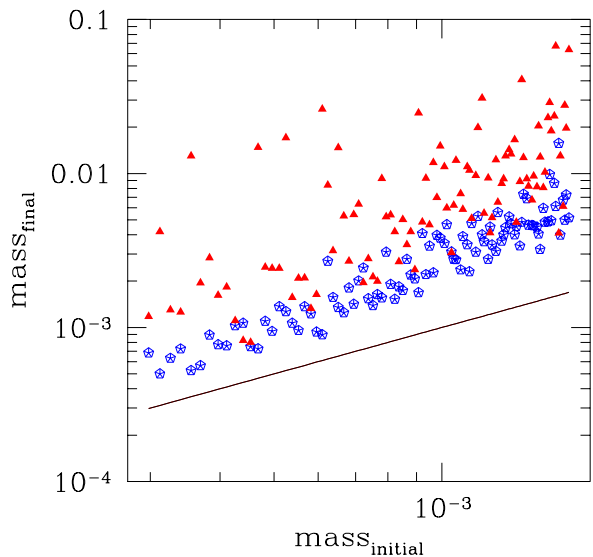,width=3.45truein,height=3.45truein}}
\caption{\label{massinfin} The intermediate ($t\approx 0.87 \tff$,
open pentagons) and final ($t\approx 1.0 \tff$, filled triangles)
masses are plotted versus the initial stellar mass for Run~C, a
cluster containing 100 stars with a distribution of initial masses.
The solid line denotes the initial range of masses. The cluster is
`cold' and contains 90 per cent of its initial mass in the form of
gas.}
\end{figure}

We have shown above that accretion can result in a spectrum of stellar
masses and in mass segregation even when the initial mass distribution
is that of equal masses. A further question is how dependent the final
masses are on the initial mass distribution. We performed a test case
with a spread of a factor of five in the initial
masses. Figure~\ref{massinfin} shows the relationship between the
initial and post-accretion stellar masses for Run C, an initially
uniform, `cold' cluster with 90 per cent of its mass initially in the
form of gas. The post-accretion mass distribution is shown when the
stars have increased their total mass (from 10 per cent) to 25 percent
and 100 per cent of the cluster mass. We see that the initial masses
play a large role in determining the masses at the intermediate stage,
ie that there is still a strong correlation between initial and post
accreted masses. On the contrary, at the later stage where the stars
have accreted all the gas, the correlation between final and initial
masses is much less clear.  There is still a general trend that the
mean final mass increases with the initial mass, but the scatter at
any given initial mass is so large (a factor 20) that a star's final
mass places only the weakest constraints on the initial mass of the
condensation from which it formed.

The reason for this difference in the intermediate and final mass
distribution lies in the cluster evolution. The cluster is initially
uniform (constant density) such that the only difference in the
accretion rates comes from the accretion radius. This is dependent on
the stellar masses (see below) and as such leads to a dependence of
the post-accretion mass on the initial mass.  The cluster is unstable
in a uniform configuration and evolves (in this case due to
gravitational collapse) to a more centrally-condensed
configuration. Once the cluster is centrally condensed, the gas
density play a more important role in the accretion rate such that the
mass accreted is more due to the gas density than due to the accretion
radius.  This is especially so if the accretion radius is only
weakly dependent on the star's mass as we shall see below. Thus, a
centrally-concentrated cluster allows even initially lower-mass stars
to have higher accretion rates, and eventually attain higher masses if they
are in regions with high gas density.

\section{An analytic prescription for competitive accretion}

In addition to seeing how accretion results in mass
segregation and a mass spectrum (Figure~\ref{massvsrad}), one of our goals is to understand
the physics of the accretion process. This involves determining what
sets the accretion rates of individual stars. A parametrised
form for the accretion rates  will 
allow the accretion process to be applied to larger groups of stars 
without needing to resolve the accretion flow around each star. 

The first possible parametrisation of the accretion rate is the
Bondi-Hoyle formalism discussed above. In this case, the accretion
radius is basically the radius where the gravitational
energy due to the star is larger than the kinetic energy (
including the unperturbed gas velocity relative to the star, $\vrel$,
and sound speed, $c_s$).
Thus (Bondi \& Hoyle~1944), 
\begin{equation}
\rbh = { 2 G\ms \over \vrel^2 + c_s^2}, 
\ee 
and
\be
\macc \approx \pi \rho \sqrt{\vrel^2 + c_s^2} \rbh^2.
\ee
Alternatively, the cluster potential sets another radius,
the tidal radius, due to the multiple gravitational
sources. This radius is analogous to the Roche-lobe radius
in accreting binary systems. We adopt this radius as
our second potential accretion radius. This tidal-lobe radius
is 
\be
\rroche = C_{\rm tidal} \left({\ms\over \menc}\right)^{1/3} \rs,
\ee
where $\ms$ is the star's mass, and $\menc$ is the mass enclosed 
within the cluster at the star's position $\rs$. In the Roche-lobe
approximation, $C_{\rm tidal} \approx 0.5$ (e.g. Pacynski~1971) which
we will adopt throughout this paper. The accretion rate is then
\be
\macc \approx \pi \rho \vrel \rroche^2.
\ee

The expectation is that the smaller of the two radii
should determine the accretion rate as it will be this
radius which ultimately decides if the gas is bound to the star
and thus will be accreted. Considering the high stellar
velocities in a cluster, the Bondi-Hoyle radius is commonly 
expected to be the smaller of the two.

\begin{figure}
\vspace{-0.35truein}
\psfig{{figure=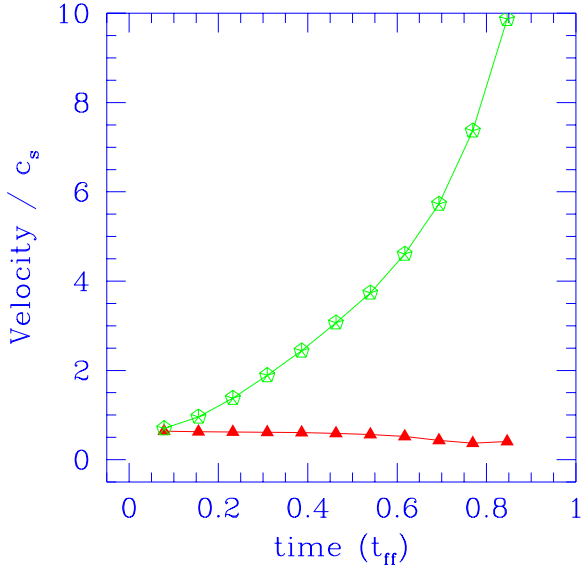,width=3.5truein,height=3.5truein}}
\caption{\label{velvstime} The stellar velocity dispersion (open
pentagons) and the mean relative gas velocity in the tidal-lobes of
the individual stars (filled triangles) are plotted in units of the
gas sound speed $c_s$ as a function of time for Run~A . The cluster
contains 100 stars, is initially uniform (stars and gas) and contains
90 per cent of its mass in gas (120 Jeans masses).  }
\end{figure}

One difficulty is estimating the relative velocity which goes into
both the Bondi-Hoyle radius (equation~10) and the Bondi-Hoyle
accretion rate (equation~11). We make two estimates of this velocity,
one calculated as the velocity of each star relative to the centre of
mass of the cluster and the second is the velocity relative to the gas
at the tidal-lobe radius. This relative velocity is calculated by
averaging each component of the three-dimensional velocity vectors
 of the particles which
overlap the tidal radius.  We shall refer to the Bondi-Hoyle accretion
estimate using this relative velocity as the {\it modified}\
Bondi-Hoyle accretion.  The evolution of each of these velocities is
shown in Figure~\ref{velvstime} for Run A, a cold cluster of 100 stars
initially comprising 10 per cent of the total mass. Both velocities
are initially subsonic but the stellar velocities quickly become
supersonic as the cluster collapses. In contrast, the relative gas
velocity at the tidal-lobe remains subsonic throughout the
evolution. This occurs as the gas is accelerated under the same
potential that the stars are. Thus, the gas tracks the stellar motions
and the relative velocity is small.

The gas density is also estimated at the tidal-lobe radius as
that is the radius at which the density is unaffected by the
star's potential. This density is calculated as the average over the particles
which overlap the tidal radius.
Estimates of the gas density and relative velocity at the
tidal-radius are generally well determined as the number of
SPH particles there is typically large. Exceptions to this
occur when a star is ejected from the cluster and the gas density
falls to zero. Another problem that was encountered was
due to the presence of other stars in the tidal-lobes. The tidal-lobe 
formulation does not allow for the effects of close binary
systems (with $R_{\rm sep} < \rroche$) so that the gas
velocities and density do not correspond to only
one star. Such systems are generally excluded from the analysis that follows
(except in \S~6.3).

\begin{figure}
\vspace{-0.35truein}
\psfig{{figure=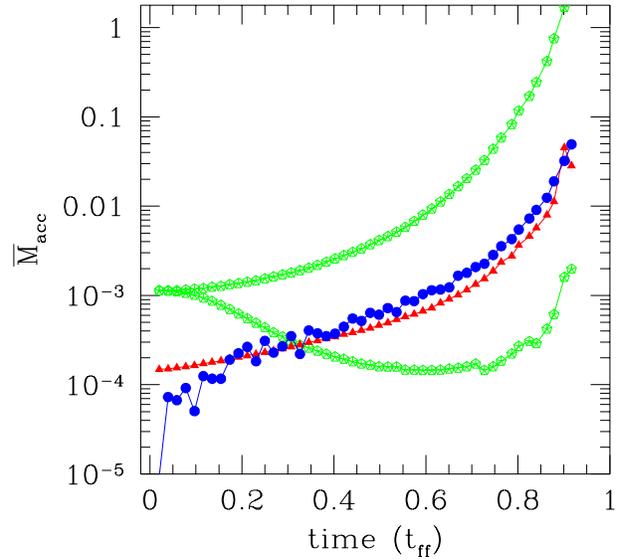,width=3.5truein,height=3.5truein}}
\caption{\label{maccavcomp} The mean accretion rate (filled circles) is
plotted as a function of time for Run A, a cluster of 100 stars which
initially has 90 per cent of its mass in gas and is cold. Estimates of
the accretion using tidal-lobe as the accretion radius (filled
triangles) and using the Bondi-Hoyle radius (open pentagons) are also
plotted. The upper Bondi-Hoyle curve uses the relative gas velocity in
each star's tidal lobe (modified Bondi-Hoyle) 
whereas the bottom curve uses each star's
velocity in the cluster.}
\end{figure}

Using the above estimates of the accretion rate as a function of the
local gas density, relative stellar velocity and the accretion radius,
we can compare these with the accretion rate determined from the SPH
simulations. Figure~\ref{maccavcomp} plots the evolution of the mean
accretion rate for Run~A, a `cold' cluster of 100 stars where 90 per
cent of the initial cluster mass is in the form of gas. Only those
stars not in binaries are included. The comparison includes two
estimates of the Bondi-Hoyle radius using each star's velocity
relative to the cluster's centre of mass and its velocity relative to
the gas in the tidal-lobe.

The mean accretion rate determined by the SPH simulations increases
with time until near the end of the simulation when the gas is 
significantly depleted. The Bondi-Hoyle accretion rates are both
initially much higher than the SPH accretion rate as the stars are 
initially at rest and the gas is cold. The modified Bondi-Hoyle accretion rate 
(using the velocity relative to the local gas) remains much higher than 
the SPH accretion rate. This is because, as shown in Figure~\ref{velvstime}, this velocity 
remains subsonic throughout the simulation, and thus $\rbh > \rroche$. 
On the other hand, the 
Bondi-Hoyle accretion rate using the stellar velocity relative to the 
cluster centre of mass decreases as the stars accelerate and move
supersonically. The accretion rate is thereafter
significantly smaller than the SPH accretion rate. In contrast, the tidal-lobe
accretion rate is generally very similar to the SPH accretion rate 
throughout the simulation. 

From this comparison we can see that the Bondi-Hoyle accretion rate 
is either too high or too low depending on which velocity is used. It 
is also readily apparent why the naive accretion rate discussed in 
Section~3 gives a  timescale that is too long.
The
stellar velocity relative to the cluster's centre of mass  is not
the appropriate velocity. It is the velocity relative to the local
gas from which the star is accreting which must be used.
Thus, from now on, we shall discuss only the 
modified Bondi-Hoyle accretion rate using the stellar velocity relative to the 
local gas.

\subsection{Single-star clusters}

\begin{figure}
\vspace{-0.35truein}
\psfig{{figure=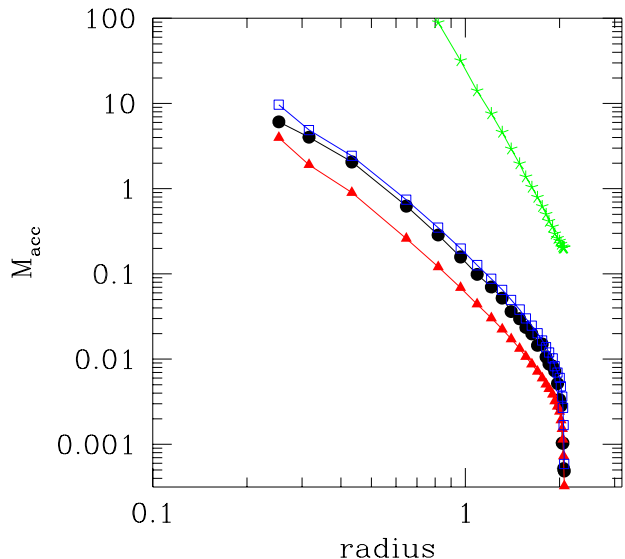,width=3.5truein,height=3.5truein}}
\caption{\label{macconestar} The accretion rate (filled circles) of a
single star in a cloud of cold gas is plotted as a function of its
radius in the cloud (Run D). The star initially contains one percent of the
total mass. Estimates of the accretion rate based on the tidal-lobe
(filled triangles) and modified Bondi-Hoyle (stars) are also plotted. The open
squares denote the tidal-lobe accretion that includes a component of
spherically-symmetric infall.}
\end{figure}

In order to ensure that we can correctly model the accretion rate, we
need to compare the different parametrisations of the accretion for
individual stars. Figure~\ref{macconestar} plots the accretion rate
onto one star in a cold gas cloud where 99 per cent of the initial
mass is in the form of gas (Run~D).  Thus, the gas provides the potential as
would do an entire cluster but without the difficulties of binary
systems and also with increased numerical resolution. The accretion
rate onto the star is plotted as a function of it's position (radius)
in the gas cloud. The accretion rate increases as the star falls in
towards the centre of the cloud, due to the increased density there as
the cloud collapses to a centrally-condensed configuration. Also
plotted in Figure~\ref{macconestar} are the modified Bondi-Hoyle
accretion rate (using the velocity relative to the gas) and two
estimates of the tidal-lobe accretion rate. The first of these
estimates includes only the relative gas velocity, $\vrel$, whereas
the second (and higher) estimate includes an adhoc estimate of the
degree of spherical infall through the tidal-lobe, $\vinf$.  This
spherical accretion can occur over $4\pi$ sterradians, but, as a first
approximation, we use and intermediate value of $3\pi$ as the
accretion is not completely spherical. Thus 
\be 
\macc = \pi \rho \racc^{2}(\vrel + 3 \vinf), 
\ee 
where $\vinf$ is the mean infall
velocity of particles at the tidal-lobe radius, $\vinf = {\bf
v}\cdot{\bf r}/|{\bf r}|$.

As seen above, the modified Bondi-Hoyle accretion rate is much larger
than the SPH accretion rate. The tidal-lobe estimates are much closer
to the numerically determined value. The estimate not including
spherical infall is slightly lower than the SPH value whereas the
estimate including the spherical accretion is very slightly higher
than the SPH value.

\begin{figure}
\vspace{-0.35truein}
\psfig{{figure=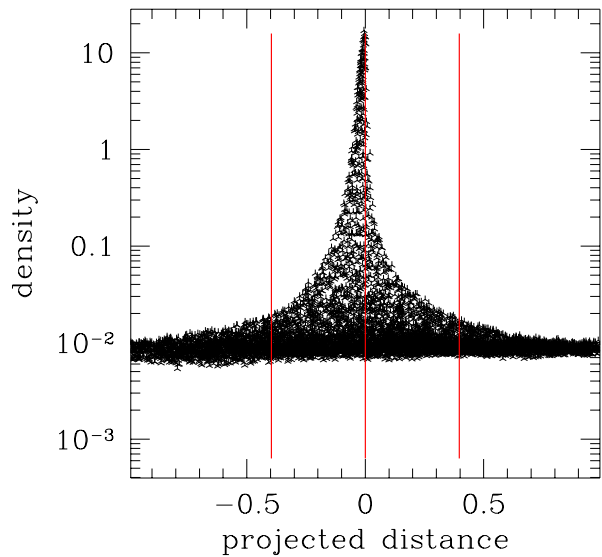,width=3.5truein,height=3.5truein}}
\caption{\label{deninlobe} The density of SPH gas particles around a
single-star cluster (Run~E) are plotted as a function of the projected
distance relative to the direction of the star's motion. The star is
travelling inwards at Mach 0.9 relative to the local gas. The size of
the tidal lobe is indicated and corresponds approximately to where the
density starts to increase towards the star. The modified Bondi-Hoyle
radius is larger than the plotted region ($\rbh \approx 2.$).  Note
that the gas density is higher at projected distances $< 0$,
indicating that the gas is predominantly accreted from behind the
star. The slight density gradient is because the star is infalling a
bit faster than the gas and thus seeing the higher density ahead.}
\end{figure}

The tidal-lobe accretion is a good fit to the accretion rate even in
simulations where the star is initially supersonic relative to the
cluster's centre of mass.  This occurs due to the convergence of the
stellar and gas velocities as the cluster collapses. Both the star's
and the gas velocities are increasingly dominated by the gravitational
acceleration of the cluster. As both the star and the gas start
collapsing towards the centre of the system, the gas velocity inside
the star's tidal-lobe becomes more similar to the star's
velocity. Thus, the relative velocity decreases and the flow becomes
subsonic. This effect is aided by the accretion as it helps converge
the star's velocity to that of the flow. Thus, $\rroche < \rbh$ and
the tidal-lobe accretion determines the accretion rate.

The finding that the tidal-lobe radius is a good approximation to the
accretion radius is further backed up by looking at the gas density
near the accreting star. Figure~\ref{deninlobe} plots the gas density
as a function of the projected distance from the star for a
single-star initially travelling at Mach~2 in a cluster potential
dominated by gas (Run~E). Figure~\ref{deninlobe} is plotted at a time by which
the gravitational acceleration of the cluster has converged the
stellar and gas motions such that both the star and the gas are moving
inwards and the relative velocity of the local gas is Mach 0.9.
The projected distance is calculated as the distance relative to the
star's velocity,
\be 
d_{\rm proj} = {{\bf r} \cdot {\bf v} \over |{\bf v}|}.
\ee 
We see that the gas density starts to increase at a distance
that corresponds roughly to the tidal-lobe radius as expected if this
is the physical accretion radius. It is also worth noting that the gas
density is somewhat skewed relative to the star's position with higher
density material at greater distances behind the star then in front of
it (relative to its motion). This occurs as generally the
material is accreted from behind the star, as in the classical
Bondi-Hoyle accretion.

\subsection{Many-star clusters}

\begin{figure}
\vspace{-0.35truein}
\psfig{{figure=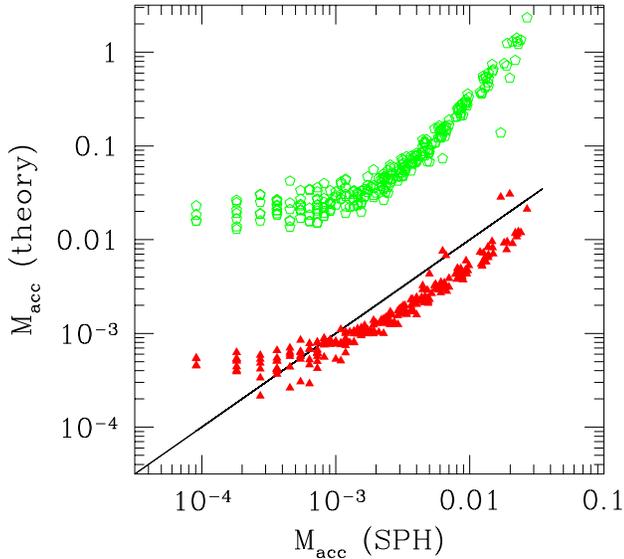,width=3.5truein,height=3.5truein}}
\caption{\label{macc30star} Estimates of the accretion rate using
the modified Bondi-Hoyle (open pentagons) and tidal-lobe (filled triangles) 
formalisms are
plotted against the SPH determined accretion rates for each star in a
cluster of 30 (Run~F). Only stars that do not have companions within $\rroche$ 
are included. The cluster is initially uniform, cold and contains 82
per cent of its mass in gas.}
\end{figure}

Accretion onto stellar clusters is considerably more complex than the
above models of a single star in a gas-dominated potential.  In a
stellar cluster, accretion is not only limited by the tidal potential
of the overall cluster but also by competitive accretion from other
stars. Figure~\ref{macc30star} plots the comparison of the SPH
accretion rate with that based upon tidal-lobe and modified
Bondi-Hoyle accretion for Run~F, a `cold' cluster of 30 stars
initially containing 82 per cent of its mass in gas. The accretion
rates are plotted for all stars that do not have another star within
their tidal-lobe. This is done as the tidal-lobe determination
neglects the presence of any companions (the effect of other stars in
the tidal-lobe is discussed in \S 6.3).

The evolution of the accretion rate for the cluster is from low
accretion rates towards higher accretion rates as the cluster
collapses under its self-gravity. The quantisation of the lowest
accretion rates is due to numerical accuracy and therefore no
conclusions are based upon these points.  We see that, as in the case
of the single-star cluster, the tidal-lobe accretion is a good
approximation to the numerical accretion rate determined by the SPH
code. In contrast, the modified Bondi-Hoyle accretion rate, using the relative
velocity inside the tidal-lobe, is much too high. It is important
to note that the tidal-lobe accretion is a good approximation to the SPH
determined accretion rate, although it generally slightly
underestimates it. In contrast, not only is the modified Bondi-Hoyle accretion
rate too high, it also diverges from the numerical accretion rate
towards the higher accretion rates. The tidal-lobe accretion is a better
fit to the accretion rate as $\rroche < \rbh$ when the local
gas velocity is used to calculate $\rbh$.

This agreement between the numerical and tidal-lobe accretion rates
was found for most clusters as long as the gas is reasonably cold.  If
the gas is sufficiently warm then $\rbh < \rroche$ and the Bondi-Hoyle
is a reasonable fit to the accretion. Such a situation occurs if the
gas contains only a few Jeans masses and thus should not have been
able to fragment to form the number of stars found in the cluster.
One possibility for a cluster with relatively warm gas is when feedback
from young stars heats up the gas. 
Other exceptions were found when many stars are in the same tidal-lobe
such as occurs when gas infalls from outside the cluster onto a
star-dominated potential, and virialised cluster (see below). 
Generally up to a third of systems have an other star within their tidal-lobes
at any given time, although this can be significantly higher in the stellar dominated potentials. In this
last case, most of the stars have $\rbh < \rroche$ although some are
still better modelled by the tidal-lobe accretion 
when their velocities are inwards.

\subsection{Accretion in a stellar dominated potential}

\begin{figure}
\vspace{-0.35truein}
\psfig{{figure=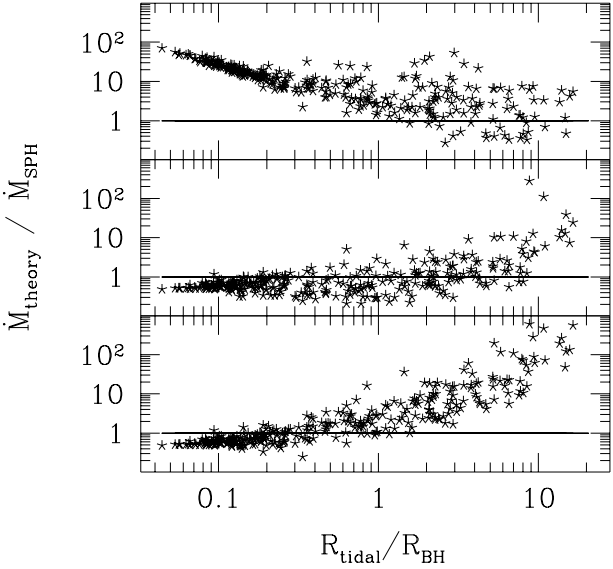,width=3.5truein,height=3.5truein}}
\caption{\label{compmacrhnh} The ratio of the theoretical
accretion rates compared to that calculated by the SPH code is plotted
against the ratio of the tidal-lobe radius to the modified Bondi-Hoyle radius
for Run~C.
The top panel uses the modified Bondi-Hoyle accretion radius whereas the bottom
panel uses the tidal-lobe radius as the accretion radius. The middle
panel uses the tidal-lobe accretion radius but adapted to include the
effects of multiple stars inside this radius (see text). The tidal-lobe radius
gives a better fit to the accretion rate when $\rroche \simless \rbh$.
The accretion rates are calculated for a cluster of 100 stars with variable
initial masses and an initial gas fraction of 90 per cent.}
\end{figure}

As noted above, companions within a star's tidal-lobe complicate the
competitive accretion picture. This becomes increasingly important
during the evolution as the gas is accreted and the stars come, at
least in the centre, to dominate the cluster potential. In such
circumstances, companion stars in each other's tidal-lobes becomes
common. In such circumstances the tidal-lobe accretion formalism
as described above does not work as the mass crossing into an individual
tidal-lobe can still be accreted by several objects. 

When a star has one or more companions inside its nominal tidal-lobe,
it's velocity can be significantly higher than the surrounding gas as
it is under the additional acceleration of the relatively close
companion(s). The extreme of this is when the gas is falling onto
a virialised cluster and hence the stellar and gas velocities are generally
uncorrelated. In this case, the modified Bondi-Hoyle radius becomes
smaller than the tidal-lobe radius ($\rbh < \rroche$) 
and thus determines the accretion rate.
Figure~\ref{compmacrhnh} plots the ratio of the theoretical to
numerical accretion rates versus the ratio of the tidal-lobe radius to
the modified Bondi-Hoyle radius for Run~C. The tidal-lobe accretion (bottom
panel) models well the accretion rate when the tidal-lobe is smaller
than the modified Bondi-Hoyle radius ($\rroche/\rbh \simless 1$). When
the tidal-lobe is significantly larger than the modified Bondi-Hoyle
radius, as is the case when other stars are present inside the
tidal-lobe, then the tidal-lobe formalism seriously overestimates the
accretion rate. The top panel of Figure~\ref{compmacrhnh} shows the
case when modified Bondi-Hoyle accretion is used.  As found above, the
modified Bondi-Hoyle formalism seriously overestimates the accretion rate when
$\rroche/\rbh \simless 1$ as the accretion is determined by the size
of the tidal-lobe. In contrast, the modified Bondi-Hoyle formalism works better
than the tidal-lobe formalism when $\rroche >> \rbh$. The scatter
towards the right of each panel in Figure~\ref{compmacrhnh} is due to
numerical resolution (low accretion rates) and sampling as the effective accretion radius
becomes small (few particles to calculate $\vrel, \rho$).

The middle panel of Figure~\ref{compmacrhnh} shows the result when tidal-lobe
accretion is adapted to include the effects of companions. This is done
by estimating the fraction of the material enterring the tidal-lobe which
is accreted by the star. In the tidal-lobe formalism, this can be calculated
by modifying equation~(13) as 
\be
\macc = \pi \rho \vinf \left({\ms^{2\over 3}\over\sum_{i=1}^{N_{\rm comp}} \msi^{2\over 3}}\right) \rroche^{2}
\ee
where 
$\vinf$ is the mean spherical infall through the tidal-lobe, and
$N_{\rm comp}$ is the number of stars in $\rroche$. The
mass  term effectively recalculates the cross-section of the tidal-lobe
 to include the companions (as $\rroche^2\propto (\ms / \menc )^{1/3}$), 
while using $\vinf$ removes the effect
of the orbital velocity as the stars move around each other inside the
tidal-lobe.
We see this estimate of the tidal-lobe accretion works fairly well
for $\rroche \simless 10 \rbh$,
and models to first order the reduction in the accretion rates
due to companions inside the tidal-lobe. The limitation of this
approach is that it neglects the individual velocities of the stars
which will modify the accretion rates. Thus, once $\rroche >> \rbh$
we are left with using the modified Bondi-Hoyle formalism.

\begin{figure}
\vspace{-0.35truein}
\psfig{{figure=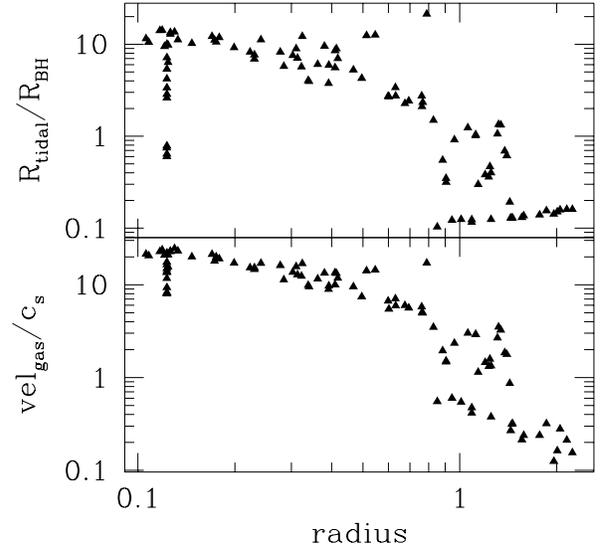,width=3.5truein,height=3.5truein}}
\caption{\label{velrgbhvsrad} The Mach number of the relative gas flow
near each star is plotted as a function of the stars' positions in the
cluster for Run~G (lower panel). The upper panel shows the
corresponding ratio of the tidal to modified Bondi-Hoyle radius as a
function of position in the cluster. The stars dominate the inner
parts of the cluster such that the stellar and gas velocities are
uncorrelated, resulting in $\rroche > \rbh$. The gas dominates the
outer parts of the cluster such that the gas and stellar velocities
are correlated and $\rroche < \rbh$. The values are averaged over
$\approx 0.1 \tff$ at $t\approx 1.0 \tff$. The relative overdensity of
stars at $r\approx 0.12$ indicates that the deepest part of the
cluster potential, where the most massive stars are, is not at the centre
of mass of the system.}
\end{figure}

As a stellar cluster accretes from the gas, the mass-fraction in the
stars increases while it decreases for the gas. Once the stars
dominate the cluster potential, their evolution is no longer
determined primarily by the collapsing gas. They then virialise and
thus have velocities uncorrelated to the infalling gas. This
transition is illustrated in Figure~\ref{velrgbhvsrad} showing the
relative gas velocity as a function of stellar position for Run~G. 
The stars are initially centrally condensed
which results in their dominating the central portions of the cluster
potential. In the outer regions of the cluster, the `cold' gas still
dominates the potential such that both stars and gas are infalling and
the relative gas velocity is subsonic.  Inside of $r=0.9$, the gas
contributes less than 25 per cent of the mass and thus no longer
determines the dynamics. The stars become virialised and the relative
gas velocities are large as the stellar and gas motions are
uncorrelated. Figure~\ref{velrgbhvsrad} also plots the ratio of the
tidal-lobe radius to the modified Bondi-Hoyle radius as a function of
position in the cluster. Stars exterior to $r=0.9$ have $\rroche
\simless \rbh$ as the relative gas velocities are subsonic whereas
stars interior to $r=0.9$ have $\rroche > \rbh$ as the relative gas
velocities are supersonic. Exceptions to this are the most massive
stars in the centre of the cluster which, due to their high mass, have
low velocities and thus $\rroche \simless \rbh$ once again.

In general as a cluster accretes from its gas reservoir, the stars
begin to dominate the potential from the inside out. This occurs due
to the higher accretion rates in the centre and due to the settling of
the accreting stars. Thus, as the gas infalls from larger radii onto
the cluster it will pass through two regimes. First, where the gas
dominates the potential, and the stellar dynamics, the accretion is
dominated by the tidal-lobes of each star. Second, further in where
the stars dominate the potential and are virialised, the accretion is
determined by a Bondi-Hoyle accretion.

\section{Discussion of accretion modelling}

In the above sections, we found that gas accretion in stellar clusters
is generally well modelled by using the tidal-lobe as the accretion
radius.  This formalism works as the cluster dynamics are dominated by
the overall collapse of the cluster due to its self-gravity. In this
case the gas in the tidal lobe of a star has a similar velocity to the
star as both are primarily due to the same gravitational
acceleration. This results in a large modified Bondi-Hoyle radius such that the
smaller tidal-lobe radius dominates the accretion,
$\racc \approx \rroche < \rbh$. Such a scenario
will always result when the gas dominates the gravitational potential
and is dynamically moving under the influence of its self-gravity. In
such a case, even if the stars are initially virialised, the changing
potential and the gas accretion will maintain similar gas and stellar
velocities.

Complications arise when more than one star is in a tidal-lobe region
as this increases the stellar velocity relative to the gas velocity in
the tidal-lobe. Gas accretion can then be modelled by including the
additional stars in the calculation of the tidal-lobe (and thus the
accretion cross section).  Once the stars dominate the gravitational
potential and are virialised, then $\rroche >> \rbh$ and accretion of
infalling gas can be well modelled with modified Bondi-Hoyle accretion
although some care has to be taken to calculate the appropriate
density and velocity of the gas.

\section{Conclusions}
We find that accretion in stellar clusters  occurs
on a dynamical timescale. Both the gas and stellar motions are governed 
by the cluster's gravitational potential. This ensures that the gas
will fill the depleted areas and thus find the stars onto which it accretes.

In clusters where gas dominates the potential, the accretion process
is better modelled by using the tidal-lobe radius, $\rroche$, as the
accretion radius rather than the commonly used Bondi-Hoyle radius,
$\rbh$. This works as both the stellar and gas velocities are
dominated by the the gas-dominated gravitational potential, and thus
have similar velocities. In this case, we have $\rroche \simless \rbh$
and the smaller $\rroche$ determines the accretion rate. This will
occur as long as the gas is the dominant component and is free to
collapse (is unsupported against gravity).  In contrast, where the
stars dominate the cluster potential and are virialised, the stellar
and gas velocities are uncorrelated resulting in high relative gas
velocities and $\rbh < \rroche$. In this case the accretion is better
modelled by Bondi-Hoyle accretion.

Accretion in a stellar cluster is highly non-uniform with
different stars accreting at significantly different rates. Coupling
this with a large mass fraction in the form of gas in the cluster
results in a spectrum of stellar masses, even from initially equal
masses. The accretion rate depends primarily on the local gas density
and thus on the star's position in the cluster as the gas quickly
concentrates towards the deepest part of the potential well. 
There is a weak
dependence on the mass ($\macc \propto \ms^{2/3}$ for tidal-lobe accretion)
such that lower initial masses bias against accretion. 
This can be overwhelmed by higher gas densities found in the
centre of the cluster. Thus, competitive accretion naturally results in a
mass-segregated cluster on the formation timescale and does not require
subsequent two-body relaxation.

\section{Acknowledgements} We thank Hans Zinnecker for useful discussions
and continual enthusiasm. 
IAB acknowledges support from a PPARC advanced fellowship.

\label{lastpage}

\end{document}